\documentstyle[aps,floats,prl]{revtex}
\input{psfig}
\begin{document}
\twocolumn

\draft

\title{\hfill LA-UR-96-1582\\
Evidence for $\bar\nu_\mu\rightarrow\bar\nu_e$
 Oscillations from the LSND Experiment at LAMPF}

\author{C. Athanassopoulos$^{12}$, L. B. Auerbach$^{12}$, R. L. Burman$^7$,\\
I. Cohen$^6$, D. O. Caldwell$^3$, B. D. Dieterle$^{10}$, J. B. Donahue$^7$,
  A. M. Eisner$^4$,\\
A. Fazely$^{11}$, F. J. Federspiel$^7$, G. T. Garvey$^7$, M. Gray$^3$,
  R. M. Gunasingha$^8$,\\
R. Imlay$^8$, K. Johnston$^{9}$, H. J. Kim$^8$, W. C. Louis$^7$,
  R. Majkic$^{12}$, J. Margulies$^{12}$,\\
K. McIlhany$^{1}$, W. Metcalf$^8$, G. B. Mills$^7$, R. A. Reeder$^{10}$,
  V. Sandberg$^7$, D. Smith$^5$,\\
I. Stancu$^{1}$, W. Strossman$^{1}$, R. Tayloe$^7$, G. J. VanDalen$^{1}$,
  W. Vernon$^{2,4}$,N. Wadia$^{8}$,\\
J. Waltz$^5$, Y-X. Wang$^4$, D. H. White$^7$, D. Works$^{12}$, Y. Xiao$^{12}$,
  S. Yellin$^3$ \\
LSND Collaboration}
\address{$^{1}$University of California, Riverside, CA 92521}
\address{$^{2}$University of California, San Diego, CA 92093}
\address{$^3$University of California, Santa Barbara, CA 93106}
\address{$^4$University of California
Intercampus Institute for Research at Particle Accelerators,
Stanford, CA 94309}
\address{$^{5}$Embry Riddle Aeronautical University, Prescott, AZ 86301}
\address{$^6$Linfield College, McMinnville, OR 97128}
\address{$^7$Los Alamos National Laboratory, Los Alamos, NM 87545}
\address{$^8$Louisiana State University, Baton Rouge, LA 70803}
\address{$^{9}$Louisiana Tech University, Ruston, LA 71272}
\address{$^{10}$University of New Mexico, Albuquerque, NM 87131}
\address{$^{11}$Southern University, Baton Rouge, LA 70813}
\address{$^{12}$Temple University, Philadelphia, PA 19122}

\date{\today}
\maketitle
\begin{abstract}A search for $\bar\nu_{\mu}\to \bar\nu_{e}$ oscillations
has been conducted at the Los Alamos Meson Physics
Facility by using $\bar\nu_\mu$ from $\mu^+$ decay at rest.
The $\bar\nu_e$ are detected via the reaction
$\bar\nu_e\,p \rightarrow e^{+}\,n$,
correlated with a $\gamma$ from $np\rightarrow d\gamma$
($2.2\,{\rm MeV}$).
The use of tight cuts to identify $e^+$ events with
correlated $\gamma$ rays
yields 22 events with $e^+$ energy between 36 and
$60\,{\rm MeV}$ and only $4.6 \pm 0.6$ background events.
A fit to the $e^+$ events between 20 and
$60\,{\rm MeV}$ yields a total excess of
$51.8^{+18.7}_{-16.9} \pm 8.0$
events. If attributed to $\bar \nu_\mu \rightarrow \bar \nu_e$
oscillations, this corresponds
to an oscillation probability of
($0.31^{+0.11}_{-0.10} \pm 0.05$)\%.
\end{abstract}

We present the results from a search for neutrino oscillations
using the Liquid Scintillator Neutrino Detector (LSND)
apparatus described in reference~\cite{bigpaper1}.
The existence of neutrino oscillations would imply that neutrinos
have mass and that there is mixing among the different flavors of neutrinos.
Candidate events in a search for the transformation
$\bar\nu_\mu\to\bar\nu_e$ from
neutrino oscillations with the LSND detector
have previously been reported~\cite{paper1} for data taken in 1993 and 1994.
Data taken in 1995 have been included in this paper,
and the analysis has been made more efficient.

Protons are accelerated by the LAMPF linac to 800 MeV kinetic energy and
pass through a series of targets, culminating with the A6 beam stop.
The primary neutrino flux comes from $\pi^+$ produced in a 30-cm-long water
target in the A6 beam stop~\cite{bigpaper1}.  The total charge
delivered to the beam stop while the detector recorded data was
1787 C in 1993, 5904 C in 1994, and 7081 C in 1995.

Most of the $\pi^+$ come to rest and decay through the sequence
$\pi^+\to\mu^+\nu_{\mu}$, followed by
$\mu^+\to e^+\nu_e\bar\nu_\mu$,
supplying $\bar\nu_\mu$ with a maximum energy of $52.8$ MeV.
The energy dependence of the $\bar\nu_\mu$ flux from decay at rest (DAR) is
very well known, and the absolute value is known to 7\%
\cite{bigpaper1,burman}.
The open space around the target is short compared to the pion decay
length, so only 3\%\ of the $\pi^+$ decay in flight (DIF).
A much smaller fraction (approximately 0.001\%) of the muons DIF,
due to the difference in lifetimes and that a $\pi^+$ must first DIF.
The total $\bar\nu_\mu$ flux averaged over the detector volume, including
contributions from upstream targets and all elements of the beam stop,
was $7.6\times 10^{-10}\bar\nu_\mu/{\rm cm}^2$/proton.

A $\bar\nu_e$ component in the beam comes from the
symmetrical decay chain starting with a $\pi^-$.  This background is
suppressed by three factors in this experiment.  First, $\pi^+$
production is about eight times the $\pi^-$ production in the beam stop.
Second, 95\%\ of $\pi^-$ will come to rest and are absorbed
before decay in the beam stop.  Third, 88\%\ of $\mu^-$ from $\pi^-$
DIF are captured from atomic orbit, a process which does not give a
$\bar\nu_e$.
Thus, the relative yield, compared to the positive channel, is estimated to
be $\sim (1/8) \times 0.05 \times 0.12 = 7.5 \times 10^{-4}$.
A detailed Monte
Carlo simulation \cite{burman}, gives a value of $7.8 \times 10^{-4}$
for the flux ratio of $\bar\nu_e$ to $\bar \nu_{\mu}$.

The detector is a tank filled with 167 metric tons of dilute liquid
scintillator, located about 30 m from the neutrino source, and
surrounded on all sides except the bottom by a liquid scintillator veto shield.
The dilute mixture allows the detection in photomultiplier tubes (PMTs) of both
\v{C}erenkov light and isotropic scintillation light, so that reconstruction
provides robust particle identification (PID) for $e^\pm$
along with the $e^\pm$ position and the direction of the event.
PID is based on the quality of the position and Cerenkov
angle fits, and on the relative amount of early light~\cite{bigpaper1}.
The detector needs to distinguish between events induced by $\bar\nu_e$
(oscillation candidates) from the events produced by the $\nu_e$.
LSND detects $\bar\nu_e$ via $\bar\nu_e p \to e^+ n,$ a process
with a well-known cross section~\cite{vogel}, followed by the
neutron-capture reaction $n p \to d\,\gamma$(2.2 MeV).
Thus the oscillation event signature consists of an ``electron'' signal,
followed by a $2.2\,{\rm MeV}$ photon correlated with the electron signal
in both position and time.  Detection of DAR $\nu_e$ in LSND is
dominated by charged current reactions on $^{12}C$, but an electron from
$\nu_e\,^{12}C \to e^-\,^{12}N$ has energy $E_e < 36$ MeV
because of the mass difference of $^{12}C$ and the lowest lying $^{12}N$
state.  Moreover, the DAR
production of a correlated photon from $\nu_e\,^{12}C \to e^- n\,^{11}N$ can
occur only for $E_e < 20$ MeV because of the threshold for free
neutron production.

Cosmic rays are suppressed at the trigger level by use of the veto shield
and by rejecting events with any evidence for a muon in the previous
15.2~$\mu$s~\cite{bigpaper1}.  Even so, the trigger rate is dominated by this
background, with actual $\nu$-induced events contributing less than
$\sim 10^{-5}$ of all triggers.  Because the data acquisition and
triggering~\cite{bigpaper1} do not depend on whether the beam is on or off, the
beam-on to beam-off duty ratio could be measured from triggered events;
it averaged 0.070 over the three years of data.  The beam-unrelated
background in any beam-on sample is thus well measured from the much
larger beam-off sample, and can be subtracted.  The cuts used to select
$e^+$ candidates are designed to discriminate heavily against this
background, so that the statistical error from this subtraction can
be kept small relative to the beam-dependent signal.

 Separation of correlated neutron-capture photons from accidental signals is
achieved using an approximate likelihood ratio, $R$~\cite{paper1,bigpaper2},
for the correlated and accidental hypotheses.  $R$ is defined using
distributions~\cite{bigpaper2} of the number of hit PMTs
for the reconstructed $\gamma$ and of the time and distance between
the primary event and that $\gamma$.  For purposes of fitting, the
$R$ distribution for accidental photons is taken from laser calibration
events.  That for correlated photons is taken from cosmic ray
neutron events either directly or as modified for the lower-energy
neutrons of interest by using a Monte Carlo simulation of the distance
distribution, with fit results averaged over the two cases.

We present analysis of the full 1993+1994+1995 data sample for two
sets of positron selection cuts.  Selection A corresponds to the criteria
used in our previous paper on the 1993 and 1994 data~\cite{paper1}.
Selection B uses new insight into the nature of the beam-off backgrounds
to further reduce these backgrounds while relaxing other criteria
to increase the signal efficiency by about 40\%.
The criteria were chosen, and efficiencies determined, using
several control samples taken as part of the data stream.
A sample of ``Michel'' electrons from the decays of
stopping cosmic ray muons is used to characterize energy calibration,
resolution and PID.  Cosmic ray neutrons stopping in the  detector are
used for the 2.2 MeV $\gamma$ properties and as a ``non-electron''
control sample for electron PID.  Other neutrino induced interactions
in the detector including $\nu_\mu\;^{12}C\rightarrow \mu^- X$~\cite{albert},
and $\nu_e\;^{12}C\rightarrow e^- X$ are also used to
check efficiencies and backgrounds.  Random triggers in
association with tank calibration are used to determine veto
efficiencies, readout deadtime, and the distribution of $R$ for
accidentally coincident $\gamma$.

The primary particle in a $\bar\nu_e$ event candidate
is required to have a PID consistent with a positron.
The selection A criteria for PID were previously described~\cite{paper1},
giving an efficiency for positrons in the $36<E_e<60$ MeV
energy range of $0.77 \pm 0.02$.  Selection B loosens the PID criteria
to increase the positron PID efficiency to $0.84 \pm 0.02$.

Selection A removed all events with the
time to the previous triggered event,
$\Delta t_p <40\,\mu{\rm s}$ to eliminate Michel electrons
from muon decay.  Selection B required $\Delta t_p$ greater than
$20\,\mu{\rm s}$, and no activities between $20\,\mu{\rm s}$ and
$34\,\mu{\rm s}$ before the event trigger time with more than 50 PMT hits
or reconstructed within 2m from the positron position.
The selection A and
B efficiencies are $0.50 \pm 0.02$ and $0.68 \pm 0.02$, respectively.
The time to any subsequent triggered event, $\Delta t_a$, is required
to be $>8 \,\mu{\rm s}$ ($\sim 4$ muon lifetimes) to remove events which
are misidentified muons which decay ($0.99 \pm 0.01$ efficiency).
The reconstructed positron location was required to be a distance $D >35$ cm
from the surface tangent to the faces of the PMTs ($0.85 \pm 0.05$ efficiency).
This assures that the positron is in a region of the tank
in which the energy and PID responses vary smoothly and are well understood.
The 35 cm cut also avoids the region of
the tank with the highest cosmic ray background.

To suppress cosmic ray neutrons, the number of associated $\gamma$
with $R>1.5$ is required to be no more than
$2$ for selection A ($0.99 \pm 0.01$ efficiency)
and no more than $1$ for selection B ($0.94 \pm 0.01$ efficiency).
Cosmic ray neutrons that enter the detector often produce one or
more additional neutrons, while recoil neutrons from the $\bar \nu_e
p \rightarrow e^+ n$ reaction are too low in energy to knock out additional
neutrons.
The number of veto shield hits associated
with the events is no more than $1$ for selection A ($0.84 \pm 0.02$
efficiency)
and no more than $3$ for selection B ($0.98 \pm
0.01$ efficiency).

Beam-off data surviving these cuts differ from the expected
neutrino interaction signal in two respects.
One is the distribution of angles between the
$e^+$ direction and its position vector relative to the tank center --
background events tend to head inwards.  The
other is in the distribution of veto hits -- cosmic ray events tend to have
more of them.  These two distributions are used in a way analogous
to the R parameter discussed earlier in defining a likelihood
ratio, $S$~\cite{bigpaper2}, which is used as the final positron selection
criterion.  For selection B, but not A, we require $S > 0.5$,
a cut that loses 13\% of the expected neutrino signal while eliminating
33\% of the beam-off background.  Including a $0.97 \pm 0.01$
data acquisition efficiency gives overall efficiencies of
$0.26\pm 0.02$ for selection A, and $0.37\pm 0.03$ for selection B.

The backgrounds to $\bar\nu_e p\to e^+ n$ followed by
$n$ capture fall into three general
classes: beam-off events (cosmic ray induced), beam-related events with
correlated neutrons,
and beam-related events with an accidental $\gamma$.
As outlined above, the cosmic ray background to beam-on events is
0.07 times the number of beam-off events which pass the same criteria.

The major sources of beam-induced backgrounds are from $\mu^-$ DAR, discussed
above, and from $\pi^+$ DIF in the beam stop.  The latter results in
a background from $\bar\nu_\mu\; p$ interactions where the final $\mu^+$
is missed, and
its Michel decay positron is mistaken for a primary $\bar\nu_e\; p$ event.
These $\bar\nu_\mu$ backgrounds are estimated using the detector Monte Carlo
simulation~\cite{bigpaper1,bigpaper2}.
The backgrounds with accidental $\gamma$ overlap
are greatly reduced by selection on the $R$ parameter.
Details of all backgrounds considered are presented in Ref.~\cite{bigpaper2}.

\begin{table}\squeezetable
\caption{The number of signal and background events
in the $36<E_e<60$ MeV energy range. Excess/Efficiency is the excess
number of events divided by the total efficiency.
The beam-off background has been scaled to the beam-on time.
B' is a restrictive geometry test.}
\label{Sig}
\begin{tabular}{lccccc}
 Selection &Signal&Beam-Off      &$\nu$ Bkgd. &Excess
&\mbox{$\displaystyle{{\rm Excess}\over{\rm Efficiency}}$} \\
\tableline
A $R \ge 0$& 221 &$133.6 \pm 3.1$&$53.5 \pm 6.8$&$33.9\pm 16.6$& $130\pm 64$\\
A $R>30$   &  13 &$ 2.8 \pm 0.4$ & $1.5 \pm 0.3$& $8.7\pm  3.6$& $146\pm 61$\\
B $R \ge 0$& 300 &$160.5 \pm 3.4$&$76.2 \pm 9.7$&$63.3\pm 20.1$& $171\pm 54$\\
B $R>30$   &  22 & $2.5 \pm 0.4$ & $2.1 \pm 0.4$&$17.4\pm  4.7$& $205\pm 54$\\
B' $R \ge 0$& 99 &$ 33.5 \pm 1.5$&$34.3 \pm 4.4$&$31.2\pm 11.0$& $187\pm 66$\\
B' $R>30$   &  6 & $0.8 \pm 0.2$ & $0.9 \pm 0.2$&$ 4.3\pm  2.5$& $110\pm 63$\\
\end{tabular}
\end{table}

\begin{figure}[htbp]
\centerline{\psfig{figure=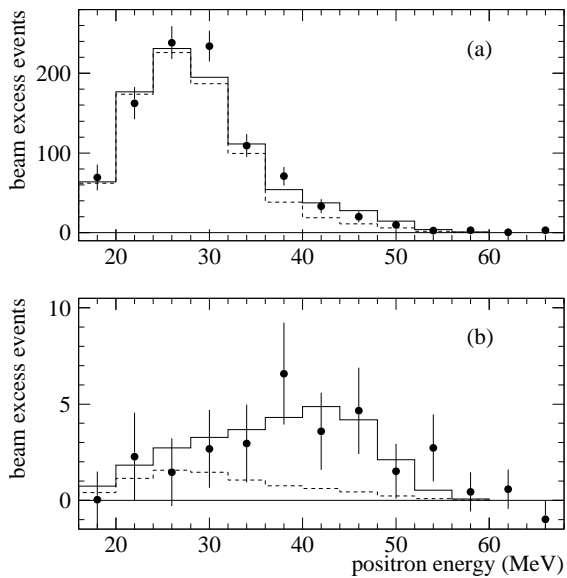,width=0.9\columnwidth,silent=}}
\caption{The energy distribution for events which pass selection B
with (a) $R \ge 0$ and
(b) $R>30$. Shown in the figure are the beam-excess data,
estimated neutrino background (dashed), and expected
distribution for neutrino oscillations at large $\Delta m^2$ plus
estimated neutrino background (solid).}
\label{evsample_e}
\end{figure}

Table~\ref{Sig} lists the number of signal, beam-off background and
neutrino-background events for the two selections with $36 < E_e < 60$ MeV --
to avoid large accidental-$\gamma$ backgrounds.
The likelihood ratio, $R$, is used to determine whether a
candidate 2.2 MeV $\gamma$ is correlated
with an electron or from an accidental coincidence.
Requiring $R>30$ (correlated-$\gamma$ efficiency = 0.23)
we observe 22 events beam-on and $36 \times 0.07 = 2.5$
events beam-off.  The estimated beam-related background consists of
$1.72\pm 0.41$ events with correlated neutrons, and $0.41\pm0.06$
without.  The probability that the beam-on events are entirely
due to a statistical fluctuation of the $4.6\pm 0.6$ event expected
total background is $4.1\times 10^{-8}$.
Figure~\ref{evsample_e}(a) shows the energy distribution of all primary
electrons which pass selection B with associated $R\ge 0$.
Figure~\ref{evsample_e}(b) shows the
electron energy distribution for selection B with $R>30$.

Kolmogorov tests have been done to check for unexpected concentrations
of events in position ({\sl e.g.}, in regions of high cosmic ray or
$\gamma$ backgrounds), energy or time (year).  No consistency
check yields a probability so low as to demonstrate a serious
inconsistency~\cite{bigpaper2}.   A restrictive geometric cut,
removing the 55\% of the selection B acceptance with highest
cosmic ray rates~\cite{bigpaper2}, also demonstrates no inconsistency;
its results are labelled B' in Table~\ref{Sig}.

\begin{figure}[htbp]
\centerline{\psfig{figure=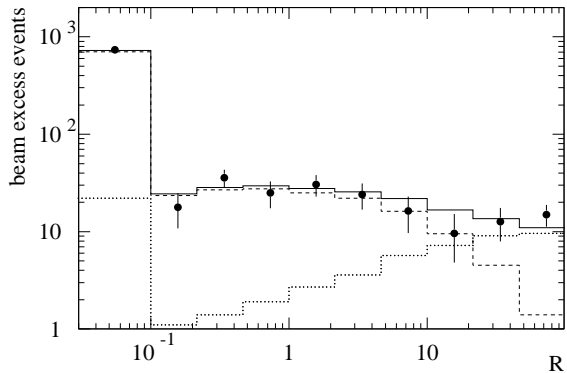,width=0.9\columnwidth,silent=}}
\caption{The R distribution, beam on minus beam-off excess, for
events that satisfy selection B and that have energies in the range
$20<E_e<60$ MeV. The solid curve is the best
fit to the data, the dashed curve is the uncorrelated $\gamma$ component of
the fit, and the dotted curve is the correlated $\gamma$ component.}
\label{rfit}
\end{figure}

To determine the oscillation probability we fit the overall $R$ distribution,
for events satisfying selection B, in the full energy range $20<E_e< 60$ MeV.
The larger energy range is used in this and the following fit to utilize the
maximum amount of data and is made possible by our increased understanding of
the background processes.
The 1763 beam-on and 11981 beam-off events were fit by a $\chi^2$
method which took spatial variations in accidental photon rates into
account by averaging the appropriate $R$ distributions at the positions
of each positron.  The result of the fit is shown in Fig.~\ref{rfit}.
It yielded $64.3^{+18.5}_{-16.7}$ beam-related
events with a correlated $\gamma$, and $860^{+17}_{-19}$ beam-related
events without a correlated $\gamma$.  The latter is consistent with a
calculated background estimate of $795\pm 134$ such events.
Subtracting the estimated neutrino
background with a correlated $\gamma$ ($12.5\pm 2.9$ events) results in
a net excess of $51.8^{+18.7}_{-16.9}$ events, corresponding to an
oscillation probability of $(0.31^{+0.11}_{-0.10}\pm 0.05)\%$, where
the second error is systematic.
A likelihood fit which uses individual local accidental-$\gamma$ $R$
distributions for each positron gave a consistent result of
$(0.27^{+0.12}_{-0.11}\pm 0.04)\%$.

\begin{figure}[htbp]
\centerline{\psfig{figure=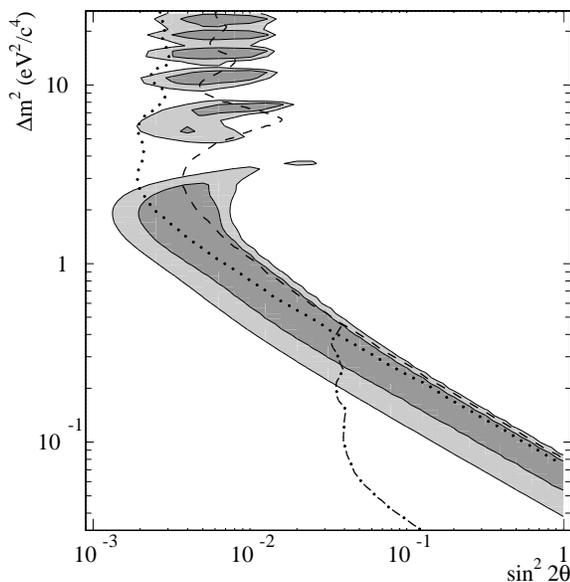,width=0.9\columnwidth,silent=}}
\caption{Plot of the LSND $\Delta m^{2}~$ $\sl vs$ $\sin^2 2 \theta~$ favored
regions.  The shaded regions are the 90\% or 99\%
likelihood regions as defined in the text, not confidence regions.
Also shown are $90\%$
C.L. limits from KARMEN at ISIS (dashed curve), E776 at BNL (dotted curve),
and the Bugey reactor experiment (dot-dashed curve).}
\label{loglik}
\end{figure}

For simplicity we present the results in the two-generation formalism,
in which the mixing probability is written as
$\displaystyle P= \sin^22\theta \sin^2(1.27 \Delta m^2 L/E_{\nu})$,
where $\theta$ is the mixing angle, $\Delta m^2$ is the difference of the
squares of the two mass eigenstates in eV$^2$, $L$ is the distance from
neutrino
production in meters, and $E_{\nu}$ is the neutrino energy in MeV.
An overall likelihood fit has been made to determine favored regions in the
$\Delta m^2$ versus $\sin^2 2\theta$ parameter space for two-neutrino
mixing.
The fit was made to distributions in the observed event energy,
the neutron likelihood ratio $R$, the reconstructed direction of the
electron relative to the neutrino beam direction,
and the distance of the primary event from the beam stop neutrino source.
The beam-related and cosmic ray backgrounds were added to the expected
neutrino oscillation signal, and a likelihood was calculated for a range
of $\Delta m^2$ versus $\sin^2 2\theta$ values.  Figure~\ref{loglik}
shows regions which are within 2.3 and 4.5 log-likelihood units of the
maximum, called 90\% or 99\%
likelihood regions.
The regions have been enlarged to account for systematic effects
by varying the inputs
to the fit to reflect uncertainty in backgrounds, neutrino fluxes
and the $R$ distribution shape.
Figure~\ref{loglik} also shows the $90\%$
C.L. limits from KARMEN at ISIS~\cite{karmen} (dashed curve),
E776 at BNL (dotted curve)~\cite{wonyong},
and the Bugey reactor experiment~\cite{bugey} (dot-dashed curve).

This paper reports the observation of 22 electron events
in the $36 <E_e < 60\,{\rm MeV}$ energy range that
are correlated in time and space with a low-energy $\gamma$ with $R>30$, and
the total estimated background from conventional processes is $4.6 \pm 0.6$
events. The probability that this excess is due to a statistical fluctuation
is $4.1 \times 10^{-8}$.
A fit to the full energy range $20<E_e<60\,{\rm MeV}$ gives an
oscillation probability of $(0.31^{+0.11}_{-0.10}\pm 0.05)$\%.
These results may be interpreted as evidence for
$\bar\nu_\mu \rightarrow \bar\nu_e$
oscillations within the favored range of Fig. \ref{loglik}.

The authors gratefully acknowledge the support of Peter Barnes,
Cyrus Hoffman, and John McClelland  during this experiment.
This work is conducted under the auspices of the US Department of Energy,
supported in part by funds provided by the University of California for
the conduct of discretionary research by Los Alamos National Laboratory.
This work is also supported by the National Science Foundation.


%
%

%


\begin{references}

\bibitem{bigpaper1}
C.~Athanassopoulos {\it et.\ al.\ }, LA-UR-96-1327, submitted to Phys.\ Rev.\
C.

\bibitem{paper1}
C.\ Athanassopoulos  {\it et.\ al.\ },
Phys.\ Rev.\ Lett. {\bf 75}, 2650 (1995).

\bibitem{burman}
R.\ L.\ Burman, M.\ E.\ Potter, and E.\ S.\ Smith,
Nucl. Instrum. Methods A{\bf 291}, 621 (1990);
R.\ L.\ Burman, A.\ C.\ Dodd, and P.\ Plischke,
Nucl. Instrum. Methods in Phys. Research A{\bf 368}, 416 (1996).

\bibitem{vogel}
C.\ H.\ Llewellyn Smith,
Physics Reports {\bf 3}, 262 (1972);
P. Vogel, Phys. Rev. D {\bf 29}, 1918 (1984);
E.\ J.\ Beise and R.\ D.\ McKeown,
Comm. Nucl. Part. Phys. {\bf 20}, 105 (1991).

\bibitem{bigpaper2}
C.~Athanassopoulos {\it et.\ al.\ }, LA-UR-96-1326, submitted to Phys.\ Rev.\
C.

\bibitem{albert}
M.\ Albert {\it et.\ al.\ },
Phys. Rev. C {\bf 51}, 1065 (1995).

\bibitem{karmen}
B.\ Bodmann {\it et.\ al.\ }, Phys.\ Lett.\ B {\bf 267}, 321 (1991);
B.\ Bodmann {\it et.\ al.\ }, Phys.\ Lett.\ B {\bf 280}, 198 (1992);
B.\ Zeitnitz {\it et.\ al.\ },
Prog. Part. Nucl. Phys., {\bf 32} 351 (1994).

\bibitem{wonyong}
L.\ Borodovsky  {\it et.\ al.\ },
Phys.\ Rev.\ Lett. {\bf 68}, 274 (1992).

\bibitem{bugey}
B.\  Achkar {\it et.\ al.\ }, Nucl. Phys. {\bf B434}, 503 (1995).

\end{references}
\end{document}